\documentclass[a4paper]{jpconf}
\usepackage{graphicx}
\begin{document}
\title{Atmosphere above a large solar pore}

\author{M Sobotka$^1$, M \v{S}vanda$^{1,2}$, J Jur\v{c}\'{a}k$^1$,
   P Heinzel$^1$ and D Del Moro$^3$}

\address{$^1$ Astronomical Institute, Academy of Sciences of the Czech Republic
         (v.v.i.), Fri\v cova 298, CZ-25165 Ond\v rejov, Czech Republic}
\address{$^2$ Charles University in Prague, Faculty of Mathematics and Physics,
          Astronomical Institute, V Hole\v{s}ovi\v{c}k\'ach 2, CZ-18000
          Prague~8, Czech Republic}
\address{$^3$ Department of Physics, University of Roma Tor Vergata, Via della
         Ricerca Scientifica 1, I-00133 Roma, Italy}

\ead{msobotka@asu.cas.cz}

\begin{abstract}
A large solar pore with a granular light bridge was observed on October 15,
2008 with the IBIS spectrometer at the Dunn Solar Telescope and a 69-min
long time series of spectral scans in the lines Ca II 854.2 nm and
Fe I 617.3 nm was obtained. The intensity and Doppler signals in the
Ca II line were separated. This line samples the middle chromosphere
in the core and the middle photosphere in the wings. Although
no indication of a penumbra is seen in the photosphere, an extended
filamentary structure, both in intensity and Doppler signals, is observed
in the Ca II line core. An analysis of morphological and dynamical
properties of the structure shows a close similarity to a superpenumbra
of a sunspot with developed penumbra. A special attention is paid to the
light bridge, which is the brightest feature in the pore seen in the
Ca~II line centre and shows an enhanced power of chromospheric oscillations
at 3--5 mHz. Although the acoustic power flux in the light bridge is five
times higher than in the ''quiet'' chromosphere, it cannot explain the
observed brightness.
\end{abstract}

\section{Introduction}
Pores are small sunspots without penumbra. The absence of a filamentary
penumbra in the photosphere has been interpreted as the indication of
a simple magnetic structure with mostly vertical field, e.g.~[1, 2].
Magnetic field lines are observed to be nearly vertical in centres of
pores and inclined by about 40$^\circ$ to 60$^\circ$ at their edges [3, 4].
Pores contain a large variety of fine bright features, such as umbral dots
and light bridges that may be signs of a convective energy transportation
mechanism.

Light bridges (hereafter LBs) are bright structures in sunspots and
pores that separate umbral cores or are embedded in the umbra. Their
structure depends on the inclination of local magnetic field and can be
granular, filamentary, or a combination of both [5]. Many observations
confirm that magnetic field in LBs is generally weaker and more inclined
with respect to the local vertical. It was shown in [6] that the
field strength increases and the inclination decreases with increasing
height. This indicates the presence of a magnetic canopy above a deeply
located field-free region that intrudes into the umbra and forms the LB.
Above a LB, [7] found a persistent brightening in the TRACE 160 nm
bandpass formed in the chromosphere. It was interpreted as a steady-state
heating possibly due to constant small-scale reconnections in the inclined
magnetic field.

In the chromosphere, large isolated sunspots are often surrounded by
a pattern of dark, nearly radial fibrils. This pattern, called
superpenumbra, is visually similar to the white-light penumbra but
extends to a much larger distance from the sunspot. Dark superpenumbral
fibrils are the locations of the strongest inverse Evershed flow -- an
inflow and downflow in the chromosphere toward the sunspot.
Time-averaged Doppler measurements indicate the maximum speed of this
flow equal to 2--3 km~s$^{-1}$ near the outer penumbral border [8].

\section{Observations and data processing}
A large solar pore NOAA 11005 was observed with the Interferometric
Bidimensional Spectrometer IBIS [9] attached to the Dunn Solar Telescope
(DST) on 15 October 2008 from 16:34 to 17:43 UT, using the DST adaptive
optics system. The slowly decaying pore was located at 25.2 N and 10.0 W
(heliocentric angle $\theta = 23^\circ$) during our observation.
According to [10], the maximum photospheric field strength was 2000 G,
the inclination of magnetic field at the edge of the pore was 40$^\circ$
and the whole field was inclined by 10$^\circ$ to the west.

The IBIS dataset consists of 80 sequences, each containing a full Stokes
($I, Q, U, V$) 21-point spectral scan of the Fe~I 617.33 nm line
(see [10]) and a 21-point $I$-scan of the Ca~II 854.2 nm line.
The wavelength distance between the spectral points of the Ca~II line
is 6.0~pm and the time needed to scan the 0.126 nm wide central part of
the line profile is 6.4 s. The exposure time for each image was set to
80~ms and each sequence took 52 seconds to complete, thus setting
the time resolution. The pixel scale of these images was 0$''$.167.
Due to the spectropolarimetric setup of IBIS, the working field of view
(FOV) was $228 \times 428$ pixels, i.e., $38''\times 71''.5$.
The detailed description of the observations and calibration procedures
can be found in [10,11].

Complementary observations were obtained with the HINODE/SOT
Spectropolarimeter [12,13]. The satellite observed the pore on
15 October 2008 at 13:20 UT, i.e., about 3 hours before the start
of our observations. From one spatial scan in the full-Stokes
profiles of the lines Fe~I 630.15 and 630.25 nm we used a part
covering the pore umbra with a granular LB.

According to [14], the inner wings ($\pm 60$ pm) of the infrared
Ca~II 854.2 nm line sample the middle photosphere at the typical height
$h \simeq 250$ km above the $\tau_{500} = 1$ level, while the centre
of this line is formed in the middle chromosphere at
$h \simeq 1200$--1400 km. This provides a good tool to study the pore
and its surroundings at different heights
in the atmosphere and to look for relations between the photospheric
and chromospheric structures.

The observations in the Ca~II line are strongly influenced
by oscillations and waves present in the chromosphere and upper
photosphere. The observed intensity fluctuations in time are caused
by real changes of intensity as well as by Doppler shifts of the line
profile.
To separate the two effects, Doppler shifts of the line profile were
measured using the double-slit method [15], consisting in
the minimisation of difference between intensities of light passing
through two fixed slits in the opposite wings of the line.
An algorithm based on this principle was applied to the time sequence
of Ca~II profiles. The distance of the
two wavelength points (``slits'') was 36~pm, so that the ``slits''
were located in the inner wings near the line core, where the intensity
gradient of the profile is at maximum and the effective formation height
in the atmosphere is approximately 1000 km. The wavelength sampling was
increased by the factor of 40 using linear interpolation, thus
obtaining the sensitivity of the Doppler velocity measurement
53 m~s$^{-1}$. The reference zero of Doppler velocity was defined
as a time- and space-average of all measurements.
This way we obtained a series of 80 Doppler velocity maps.
This method does not take into account the asymmetry of the line
profile, e.g., the changes of Doppler velocity with height in the
atmosphere.
Using the information about the Doppler shifts, all Ca~II
profiles were shifted to a uniform position with a subpixel accuracy.
This way we obtained an intensity data cube ($x, y, \lambda$, scan),
where, if we neglect line asymmetries and inaccuracies of the method,
the observed intensity fluctuations correspond to the real intensity
changes. The oscillations and waves were separated from the slowly
evolving intensity and Doppler structures by means of the 3D $k-\omega$
subsonic filter with the phase-velocity cutoff at 6 km s$^{-1}$.

\section{Results}
Examples of intensity maps in the continuum 621 nm, blue wing and
centre of the line Ca~II 854.2 nm and a Doppler map are shown in
Fig.~1. Oscillations and waves are filtered out from these images.
A~filamentary structure around the pore, composed of radially oriented
bright and dark fibrils, is clearly seen in the line-centre intensity
and Doppler maps. The area and shape of the filamentary structure
are identical in all pairs of the line-centre intensity and Doppler images
in the time series. The fibrils begin immediately at the umbral border
and in many cases continue till the border of the filamentary structure.
Their lengths are identical in the Doppler and intensity maps.
However, the fibrils seen in the line centre are spatially uncorrelated
with those in the Doppler maps.

\begin{figure}[t]
\begin{minipage}{11cm}
\includegraphics[width=11cm]{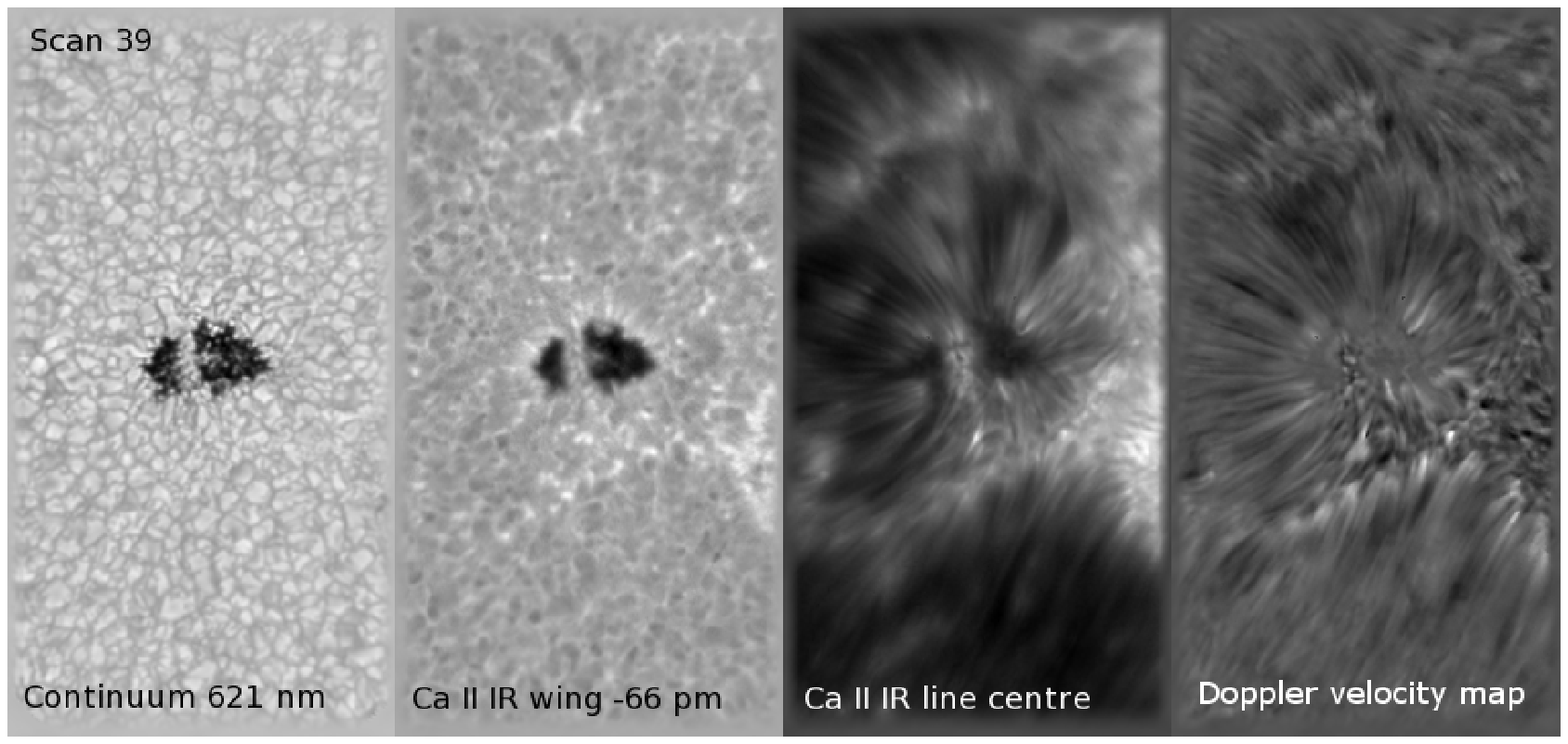}
\caption{\label{fig1}Pore NOAA 11005, full FOV $38''\times 71''.5$.
North is on the top, west on the left. From left to right: continuum 621 nm,
Ca~II 854.2 nm blue wing intensity, line centre intensity, Doppler map with
velocity range from $-2.3$ km s$^{-1}$ (black, toward the observer) to
4.7 km s$^{-1}$ (white, away from the observer).}
\end{minipage}\hspace{0.5cm}
\begin{minipage}{4.5cm}
\begin{center}
\includegraphics[width=3.3cm]{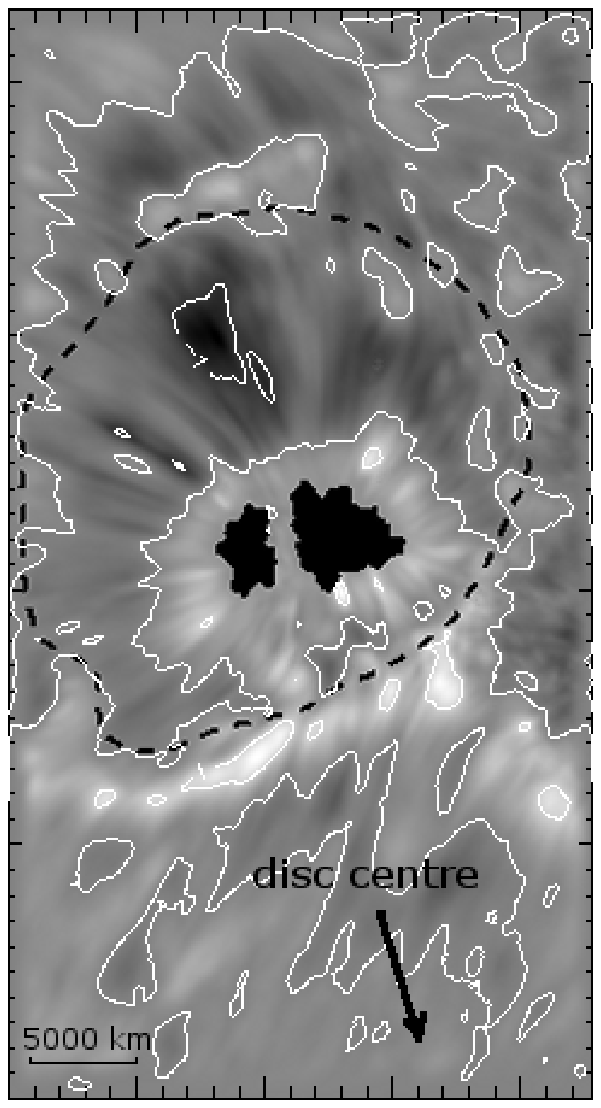}
\caption{\label{fig2}Time-averaged Doppler map with contours of zero, $-1$ and 1 km~s$^{-1}$.}
\end{center}
\end{minipage}
\end{figure}

Concentric running waves originating in the centre of the pore are observed
in the unfiltered time series of the line-centre intensity and Doppler
images. They propagate through the filamentary structure to the distance
3000--5000~km from the visible border of the pore with a typical speed
of 10 km~s$^{-1}$. These waves are very similar to running penumbral
waves observed in the penumbral chromosphere and in superpenumbrae
of developed sunspots [16].

The subsonic-filtered time series of Doppler maps was averaged in time
to obtain a spatial distribution of mean line-of-sight (LOS) velocities
around the pore. The result is shown in Fig.~2 together with contours
of zero, $-1$ and 1 km~s$^{-1}$. We can see from the figure that the
inner part of the filamentary structure contains a positive LOS velocity
(away from us,  a downflow), while the outer part, located mostly on the
limb side, shows a negative LOS velocity (toward us, an upflow). Taking
into account the heliocentric angle $\theta = 23^\circ$ and the fact
that the plasma moves along magnetic field lines forming a funnel,
the negative LOS velocity is only partly caused by real upflows but
mostly by horizontal inflows into the pore. The picture of plasma
moving toward the pore and flowing down in its vicinity is consistent
with the inverse Evershed effect observed in the sunspots' superpenumbra.

All these facts are leading to the conclusion that the filamentary
structure observed in the chromosphere above the pore is equivalent
to a superpenumbra of a developed sunspot. The missing correlation
between the fibrils seen in the line centre and those in the Doppler
maps further supports this conclusion, because, according to [17],
flow channels of the inverse Evershed effect are not identical with
superpenumbral filaments.\\

A strong granular LB separates the two umbral cores of the pore.
It is the brightest feature inside the pore in the photosphere as well
as in the chromosphere, where it is by factor of 1.3 brighter than the
average brightness in the FOV. The granular structure of the LB is
preserved at all heights, from the photospheric continuum level to
the formation height of the Ca~II line centre. It is interesting that
while a typical pattern of reverse granulation, observed in the Ca~II
wings, appears outside the pore in the middle photosphere
($h \simeq 250$ km), the LB is always composed of small bright granules
separated by dark intergranular lanes (Fig.~1). A feature-tracking
technique was applied to correlate the LB granules in position and time
at different heights in the atmosphere. A correlation was found between
the photospheric LB granules in the continuum and Ca II wings (correlation
coefficient 0.46). On the other hand, there is no correlation between
the chromospheric LB ``granules'' observed in the Ca~II line centre
and the photospheric ones in the wings and continuum.

The feature-tracking technique has shown that the mean size of the
LB granules increases with height from 0$''$.45 in the continuum to
0$''$.50 in the Ca~II wings and 0$''$.54 in the Ca~II line centre. Similarly,
the average width of the LB increases with height from 2$''$ (continuum)
to 2$''$.5 (Ca~II wings) and 3$''$ (Ca~II line centre). On the other hand,
it can be expected that the width of the LB magnetic structure decreases
with height due to the presence of magnetic canopy. The complementary
HINODE observations in two spectral lines Fe~I 630.15 and 630.25 nm
made it possible to obtain vertical stratification of temperature and
magnetic field strength in the LB photosphere, using the inversion
code SIR [18]. The results, summarised in Table~1, show that the
width of LB in temperature maps really increases with height, while
the width of the LB magnetic structure decreases and the magnetic
field strength increases, confirming the magnetic canopy configuration.

\begin{table}[h]
\caption{\label{tab1}
Widths of the light bridge measured in temperature and magnetic field
at different geometrical heights in the photosphere. $B_{\rm min}$ is
the minimum magnetic field strength in the light bridge.}
\begin{center}
\begin{tabular}{rrrr}
\br
Height (km) & Width in $T$ & Width in $B$ & $B_{\rm min}$ (G)\\
\mr
0   & 2$''$.7 & 1$''$.7 & 0\\
90  & 2$''$.7 & 1$''$.5 & 100\\
180 & 2$''$.7 & 1$''$.3 & 300\\
270 & 2$''$.9 & 1$''$.1 & 500\\
350 & 3$''$.3 & 0$''$.8 & 700\\
\br
\end{tabular}
\end{center}
\end{table}

Power spectra of chromospheric oscillations were calculated using
the unfiltered time series of the Ca~II line-centre intensity and Doppler
images. Power maps derived from the Doppler velocities at frequencies
3--6 mHz are shown in Fig.~3. At the LB position, we can see a strongly
enhanced power around 3--5 mHz, comparable with that in a plage near
the eastern border of the pore A similar power enhancement was reported
in [19]. Usually, low-frequency oscillations do not propagate through the
temperature minimum from the photosphere to the chromosphere due to the
acoustic cut-off at 5.3 mHz [20]. To explain our observations, we assume
that the low-frequency oscillations leak into the chromosphere along an
inclined magnetic field [19,21], which is present in the LB and
in the plage. The inclined magnetic field in the LB is verified by
inversions of the full-Stokes Fe~I 617.33 nm profiles [10]. The
inclination angle, extrapolated to the height of the temperature minimum,
is 40$^\circ$--50$^\circ$ to the west thanks to inclined magnetic field
lines at the periphery of the larger (eastern) umbral core. These
field lines pass above the magnetic canopy of the LB.

\begin{figure}[t]
\begin{center}
\includegraphics[width=16cm]{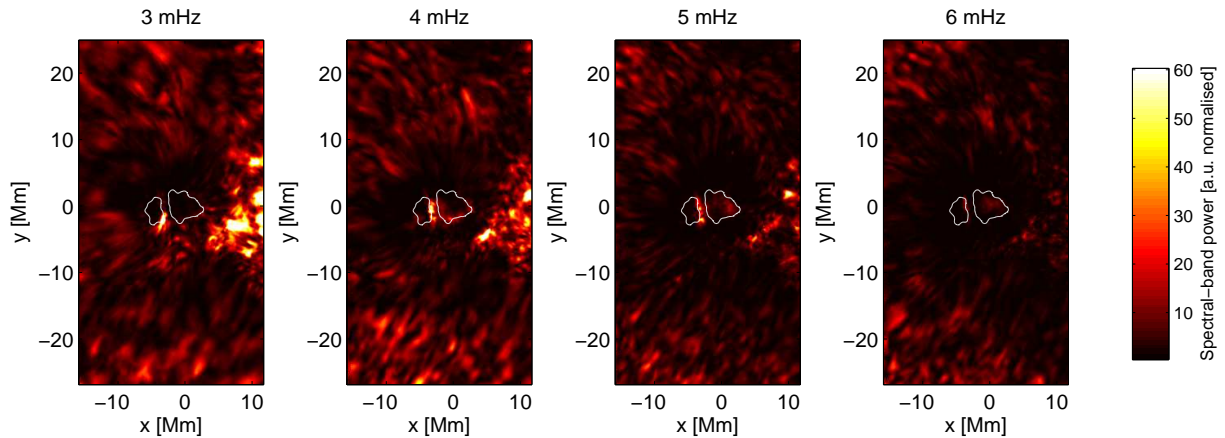}
\caption{\label{fig3}Power maps of Doppler velocities at frequencies
3--6 mHz. The contours outline the boundaries of the pore and light
bridge, observed in the continuum.}
\end{center}
\end{figure}

Following [22], we estimate the acoustic energy flux in the LB
chromosphere and compare it with the flux in the ``quiet'' region.
The observed Doppler velocities are used for this purpose. With
the estimated magnetic field inclination of 50$^\circ$, the effective
acoustic cut-off frequency decreases to the value of 3 mHz in the LB.
The total calculated acoustic power flux is 550~W~m$^{-2}$ in the LB,
while only 110~W~m$^{-2}$ in the ``quiet'' chromosphere. These results
seem to be lower than 1840 W~m$^{-2}$ presented in [22] but one has
to bear in mind that this value was obtained for the photosphere,
where the power of the acoustic oscillations must be higher than in
the chromosphere.

\section{Discussion and conclusions}
We studied the photosphere and chromosphere above a large solar pore
with a granular LB using spectroscopic observations with spatial resolution
of 0$''$.3--0$''$.4 in the line Ca~II 854.2 nm and photospheric Fe~I lines.
We have shown that in the chromospheric filamentary structure around
the pore, observed in the Ca~II line core and Doppler maps, the
inverse Evershed effect and running waves are present. Chromospheric
fibrils seen in the intensity and Doppler maps are spatially uncorrelated.
From these characteristics and from the morphological similarity of the
filamentary structure to superpenumbrae of developed sunspots we
conclude that the observed pore has a kind of a superpenumbra, in
spite of a missing penumbra in the photosphere.

A special attention was paid to the granular LB that separated the pore
into two umbral cores. The magnetic canopy structure [6] is confirmed
in this LB. In the middle photosphere ($h \simeq 250$ km, Ca~II wings),
the reverse granulation is seen around the pore but not in the LB. The
reverse granulation is explained by adiabatic cooling
of expanding gas in granules, which is only partially cancelled by
radiative heating [23]. In the LB, hot (magneto)convective plumes at
the bottom of the photosphere cannot expand adiabatically in higher
photospheric layers due to the presence of magnetic field and the
radiative heating dominates, forming small bright granules separated
by dark lanes. The positive correlation between the LB structures in
the continuum and Ca~II line wings indicates that the middle-photosphere
structures are heated by radiation from the low photosphere.
Since the mean free photon path in the photosphere is larger than 1$''$
for $h > 120$ km, the LB observed in the line wings is broader and
its granules are larger than in the continuum due to the diffusion
of radiation.

In the middle chromosphere ($h \simeq 1300$ km, Ca~II centre), the LB
is the brightest feature in the pore and it is brighter by factor
of 1.3 than the average intensity in the FOV. Since the height in the
atmosphere is well above the temperature minimum, the radiative heating
cannot be expected. The heating by acoustic waves seems to be a candidate,
because the acoustic power flux in the LB is five times higher than in
the ``quiet'' chromosphere. To check this possibility, we have to compare
the acoustic power flux with the total radiative cooling. An average
profile of the Ca~II line in the LB was used to derive a simple
semi-empirical model based on the VAL3C chromosphere [24], with the
temperature increased by 3000 K in the upper chromospheric layers
($h > 900$ km). The net radiative cooling rates
were calculated using this model. The resulting height-integrated radiative
cooling is approximately 6700~W~m$^{-2}$ in the LB chromosphere and
3000~W~m$^{-2}$ in the ``quiet'' chromosphere. The acoustic power fluxes
(550~W~m$^{-2}$ and 110~W~m$^{-2}$, respectively) are by an order of
magnitude lower than the estimated total radiative cooling, so that
the acoustic power flux does not seem to provide enough energy to reach
the observed LB brightness.

\ack{This work was supported by the Czech Science Foundation under grants
P209/12/0287 and P209/12/P568 and by the project RVO:67985815
of the Academy of Sciences of the Czech Republic.
The Dunn Solar Telescope is located at the National Solar Observatory,
operated by AURA for the National Sience Foundation.
HINODE is a Japanese mission developed and launched by ISAS/JAXA, with
NAOJ as domestic partner and NASA and STFC (UK) as international partners.
It is operated by these agencies in cooperation with ESA and NSC (Norway).}

\end{document}